\documentstyle[epsf,prl,aps,floats]{revtex}

\begin{document}
\draft
\wideabs{ 

\title{{\em Ab Initio} Calculation of Spin Gap Behavior in
CaV$_4$O$_9$}
\author{C. Stephen Hellberg,$^1$ W. E. Pickett,$^{1,2}$ L. L. Boyer,$^1$
Harold T. Stokes,$^3$ and Michael J. Mehl$^1$}
\address{$^1$Center for Computational Materials Science,
Naval Research Laboratory, Washington DC 20375}
\address{$^2$Department of Physics, University of California, Davis CA 95616}
\address{$^3$Department of Physics and Astronomy,
Brigham Young University, Provo UT 84602}
\date{\today}
\maketitle
\begin{abstract}
Second neighbor dominated exchange coupling in CaV$_4$O$_9$ has been
obtained from {\it ab initio} density functional (DF) calculations.
A DF-based self-consistent atomic deformation model reveals that the
nearest neighbor coupling is small due to strong cancellation among
the various
superexchange processes.  
Exact diagonalization of the predicted Heisenberg
model yields spin-gap behavior in
good agreement with experiment.
The model is refined by fitting to the experimental susceptibility.
The resulting model agrees very well with the experimental susceptibility
{\em and} triplet dispersion.
\end{abstract}

}

CaV$_4$O$_9$ was the first two-dimensional system observed to
enter a low-temperature quantum-disordered phase
with a spin gap $\Delta \approx 110$K.
The gap
was first apparent in its susceptibility, which vanishes at low temperatures as 
$\chi(T \rightarrow 0) \sim \exp(-\Delta/kT)$
\cite{taniguchi95a}, and was observed directly in
the dispersion of triplet spin excitations ($\Omega_Q$) measured
by neutron scattering\cite{kodama97a}.  This unexpected behavior
has stimulated considerable theoretical study
of the exchange couplings between S=$\frac{1}{2}$ spins on the V lattice
using Heisenberg models
\cite{gelfand96a,fukumoto98a,weihong98a,theory}.

\begin{figure}[t]
\epsfxsize=3.000in\centerline{\epsffile{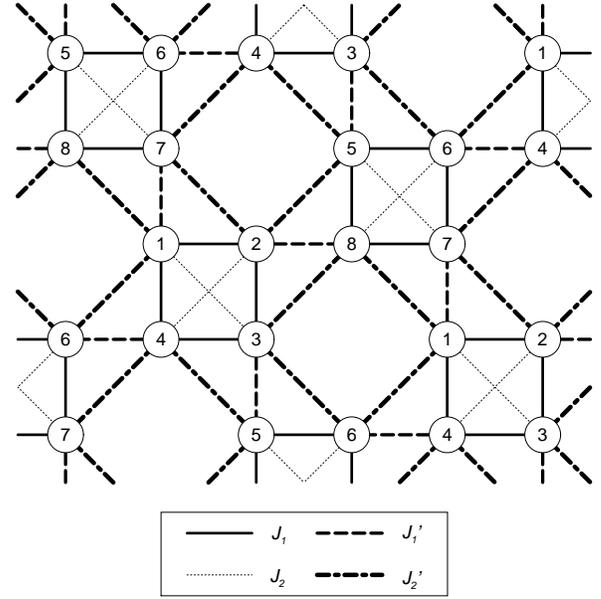}}
\vspace{.1in}
\caption{
Couplings in CaV$_4$O$_9$.  The circles represent V atoms, and the lines
between V's show the couplings.
The numbers label the sites used in the LSDA calculations
to determine the couplings.
Line thicknesses are proportional
to the best set of couplings found by fitting to the experimental susceptibility.
The strongest coupling,
$J_2^{\prime}$, is shown as the thick dot-dashed lines, forming metaplaquettes,
e.g., 1-6-3-8.
}
\label{fig:lattice}
\end{figure}

CaV$_4$O$_9$ is a layered compound---the interlayer distance
is sufficiently large to make interlayer V-V coupling negligible.
Within a layer, the V atoms form a
\mbox{$\frac{1}{5}$-depleted} square lattice
shown as the circles in Fig.\ \ref{fig:lattice} \cite{bouloux73a,pickett97a}.
The lattice was originally viewed as an array of 
square ``plaquettes" of V ions (e.g., 1-2-3-4 in Fig.\ \ref{fig:lattice})
tending toward singlet formation
since isolated plaquettes have a singlet ground state.  
Examination of the structure however suggests intra- and inter-plaquette
nearest neighbor V-V coupling
should be similar,
so the limit of isolated plaquettes is not realistic.

Self-consistent electronic
structure work\cite{pickett97a} identified the 
V$^{4+}$ spin orbital as $d_{xy}$, which
implied that it was a larger square of V ions, the ``metaplaquette,''
where singlet formation arises.  Fitting Heisenberg Hamiltonians
to the measured dispersion of the triplet excitations confirmed that
the dominant second neighbor exchange coupling is crucial to account
for the shape of $\Omega_Q$\cite{kodama97a}. 

The complete Heisenberg Hamiltonian for CaV$_4$O$_9$ has four different
coupling constants:
nearest-neighbor ($nn$) and next-nearest-neighbor ($nnn$) couplings
and, for each of these, intra- and inter-plaquette couplings.
In notation of Gelfand {\it et al.}\cite{gelfand96a},
the Hamiltonian is given by
\begin{eqnarray}
H & = & J_1 \sum_{nn}   {\bf S}_i \cdot {\bf S}_j 
    +   J'_1\sum_{nn'}  {\bf S}_i \cdot {\bf S}_j \nonumber
\\
  & + & J_2 \sum_{nnn}  {\bf S}_i \cdot {\bf S}_j 
    +   J'_2\sum_{nnn'} {\bf S}_i \cdot {\bf S}_j,
\label{eq:hamiltonian}
\end{eqnarray}
where ${\bf S}_i$ denotes the spin $\frac{1}{2}$ operator in site $i$.
The $nn$ sums run over nearest-neighbor bonds and
the $nnn$ sums run over next-nearest-neighbor bonds.
Unprimed sums connect V's in the same plaquette,
while primed sums connect V's in different plaquettes.
The four couplings are drawn in different line styles in
Fig.\ \ref{fig:lattice}.

In this Letter we show that the spin gap behavior of CaV$_4$O$_9$, even
considering its complex structure with eight very low symmetry V$^{4+}$
ions in the primitive cell, can
be calculated in {\it ab initio} fashion.  Our work has three separate
aspects.
1) Local spin density approximation (LSDA) calculations are used to obtain
energies for various magnetic configurations.
The resultant exchange interactions
are obtained by fitting these energies to the mean-field Heisenberg model as 
described below.
2) An approximate but physically motivated local orbital
method 
called the self-consistent atomic deformation (SCAD) method
\cite{scad}
is used to provide explicit local orbitals,
eigenvalues, and hopping integrals
for calculating the exchange interactions from
perturbation theory.  
This method reveals that 
the $nn$ interactions are not intrinsically small, but the
net value of the superexchange coupling
is small due to cancellations among various fourth-order
processes.
It also indicates that {\it direct} V-V exchange coupling is 
important.
3) The Heisenberg Hamiltonian is solved using exact
diagonalization techniques on finite periodic clusters.  Spin gap behavior
is obtained, and
$\chi$(T) is similar to the data.  
The Heisenberg couplings are refined by fitting to $\chi(T)$.
The resulting Hamiltonian agrees well with $\chi(T)$ and with the
triplet dispersion determined from neutron scattering.

The LSDA calculations of the energy for various magnetic configurations
were more precise extensions of previous work on CaV$_4$O$_9$
\cite{pickett97a,lapw}.
The magnetism of the V ion is found to be robust, allowing us to break
the spin symmetry in any manner we choose and obtain the energy from
a self-consistent calculation.  
The symmetry of the non-magnetic state is initially broken
as desired by applying
the necessary local magnetic fields to the V ions.
The seven configurations we have 
chosen include the ferromagnetic (FM) state, 
one ferrimagnetic (FiM) state, and 
five antiferromagnetic (AF) states with zero net spin.  
These AF states include the N\'eel state,
a state in which FM plaquettes are antialigned (FMPL),
and
a state in which the metaplaquettes are aligned
antiferromagnetically (AFMP).
The configurations, given explicitly in Table \ref{tab:configurations},
were chosen either because of their
physical relevance (AFMP was anticipated to be lowest in energy, as found)
or computational considerations such as retaining inversion symmetry.

\begin{table}[tb]
\caption[]{Magnetic configurations of the eight V ions in the primitive
cell for the states used to determine the exchange constants
from LSDA.
Most configurations are defined in the text.
V ions are numbered as in Fig.\ 1.
The final column shows the relative LSDA energies.
}
\begin{tabular}{ccccccccccr}
      & 1  & 2  & 3  & 4  &   & 5  & 6  & 7  & 8  &$\Delta E/8$\\
\tableline
FM    & +  & +  & +  & +  &   & +  & +  & +  & +  &   0.0 meV\\
FMPL  & +  & +  & +  & +  &   & -- & -- & -- & -- & -95.2 meV\\
FiM   & +  & +  & +  & -- &   & +  & -- & +  & +  &   -70.7  meV\\
AFMP  & +  & -- & +  & -- &   & +  & -- & +  & -- &-130.6 meV\\
N\'eel& +  & -- & +  & -- &   & -- & +  & -- & +  & -35.4 meV\\
STEP  & +  & -- & -- & +  &   & -- & +  & +  & -- & -74.8 meV\\
STEP2 & -- & +  & +  & -- &   & -- & +  & +  & -- & -81.6 meV
\end{tabular}
\label{tab:configurations}
\end{table}

The resulting energies were fit to the mean-field Heisenberg model,
which contains simply the ${\rm S}_i^z$ or Ising terms of the
full Hamiltonian (\ref{eq:hamiltonian}),
to determine the four coupling constants.
The six energy differences lead to six conditions on the four $J$s,
and a least-squares fit gives the values listed
as LSDA in Table \ref{tab:allJs},
each with a fitting uncertainty of about 1 meV.
Since both nearest and next nearest couplings are
AF in sign, there is a great deal of frustration in the magnetic system.  
The large value of $J_2^{\prime}$ indicates that singlet formation on the
metaplaquette is the driving force for the spin gap.

\begin{table}[t]
\caption{Values for the four couplings (in meV).
The LSDA values are derived from the energies in Table I.
The SCAD results are derived from the local orbital method,
and the Fit results come from fitting the experimental susceptibility.
Both are described later in the paper.
Also shown are the couplings deduced from neutron scattering
data \protect\cite{fukumoto98a,weihong98a}.
}
\begin{tabular}{lcccr}
Method  & $J_1$ & $J_1^{\prime}$ & $J_2$ & $J_2^{\prime}$ \\
\tableline
LSDA  & 8.9 & 1.1 & 6.5 & 23.8  \\
SCAD  & 9.7 & 12.5 & 3.9 & 19.3  \\
Fit   & 9.3 & 9.6 & 3.7 & 14.2  \\
Neutron & 6.8 & 6.8 & 1.7 & 14.0  \\
\end{tabular}
\label{tab:allJs}
\end{table}

To understand how these values of the exchange parameters arise, we 
evaluate the fourth-order expressions for the exchange constants, 
using an approximate but parameter-free method
based on the 
SCAD
method.
For
each coupling constant in CaV$_4$O$_9$,
we focus on the relevant clusters for each coupling.
The
$nnn$
interactions require a V$_2$O cluster
with two V ions (each with one relevant orbital)
and one O in between.
The
$nn$
exchange interactions require a V$_2$O$_2$ cluster.
All three $2p$ orbitals in each O are relevant,
since the low symmetry makes them non-degenerate and oriented in
directions determined not by symmetry but by electronic interactions.

We neglect the Hubbard $U$ and Hund's rule coupling on the O ions.
In what follows, $U$ is the V on-site repulsion, 
$\epsilon_V$ and $\epsilon_\alpha$ are site energies of the V
and $\alpha$-th O orbitals, 
and $t_{i\alpha}$ is the
hopping amplitude
between the $i$-th V and the $\alpha$-th O orbital.
Defining the energy denominators
$\Delta_\alpha = U + \epsilon_V - \epsilon_\alpha$ 
simplifies
the
expressions.

The initial state has each O orbital doubly filled
and each V with one electron.
The perturbation theory is given by three fourth-order terms
and the direct second-order V-V term:
\begin{eqnarray}
J & = & j_1 + j_2 + j_3 + j_d \nonumber \\
  & = &
 \frac{4}{U} \left( \sum_\alpha \frac{t_{1\alpha}
 t_{2\alpha}}{\Delta_\alpha}\right)^2
+
 4 \sum_\alpha \frac{(t_{1\alpha}
 t_{2\alpha})^2}{\Delta_{\alpha}^{3}}
\nonumber \\
& + & 
   4 \sum_{\alpha<\beta}
      \frac{t_{1\alpha} t_{2\alpha} t_{1\beta} t_{2\beta} }
      {\Delta_\alpha + \Delta_\beta}
      \left(
         \frac{1}{\Delta_\alpha} + \frac{1}{\Delta_\beta}
      \right)^2
 + \frac{4t_{12}^2}{U}
\label{eq:perturb}
\end{eqnarray}
In the
$nnn$
case, $\alpha$ and $\beta$ sum over
the three orbitals in the single oxygen atom.
In the
$nn$
case, $\alpha$ and $\beta$ sum over
the six orbitals in both oxygen atoms.
The first three terms in (\ref{eq:perturb})
can be categorized by their configurations after the
second hop of the four-hop process:
1) One vanadium empty;
2) One oxygen orbital empty;
3) Two oxygen orbitals half filled.
The last term has an extra factor of two because it arises twice:
the total spin singlet case is reduced in energy
and the total spin triplet is increased by the same amount.
The latter picks up a minus sign due to electron exchange.

This expression is evaluated with the 
SCAD
model, which expresses the total 
density $n(r)$ as a sum over
localized densities $|{\phi_{\alpha}^{(i)}}({\bf r}-{\bf R}_i)|^2$
centered at the atomic sites ${\bf R}_i$ \cite{scad}.
The orbitals  $\phi_{\alpha}^{(i)}$ are solutions to atom-centered
one-electron
Hamiltonians $H_i$ for each site.
The potentials in $H_i$ are determined self-consistently
from the expression for the functional
derivative of the total energy.
It includes a local approximation
for exchange and correlation energy\cite{hedin71a} and the Thomas-Fermi 
function for kinetic energy of overlapping
densities.

Each V ion has
the lowest of its five 3d levels occupied by a single electron, giving
the V$^{4+}$, O$^{2-}$ ionic description.
$U \approx 3.5$ eV
was computed by minimizing the SCAD energy subject to the constraint
that one V ion has its charge increased by unity.
The electron
comes mainly from the other V ions
with only a minor portion coming from the nearby O ions.

The matrix elements, $t_{ij}= \langle{\psi_i}|H|{\psi_j} \rangle$
require the full Hamiltonian $H$ and orthogonalized orbitals $\psi$.
The $\psi$'s are obtained from the SCAD orbitals using L\"{o}wdin's
method \cite{lowdin50a}, and $H$ is determined from the site centered 
SCAD Hamiltonians by removing the kinetic energy overlap contributions
from the latter.
This gives expressions for $H$ that differ in the
site selected for spherical harmonic expansion of the potential. 
We find the two possibilities, $t_{ij}$ and $t_{ji}$, may differ
by $\sim$20\%, which leads to a much larger uncertainty in the
fourth-order $J$'s.
Since the vanadium sites of a given pair of V ions 
are equivalent by symmetry, the direct interaction, 
$j_d$,
has no such
uncertainty.
To be consistent with the direct interaction calculation,
we use the vanadium-site-expanded potentials for evaluating
matrix elements between oxygen-vanadium pairs.
The net values obtained
(labelled SCAD in Table \ref{tab:allJs})
agree rather well with those derived from LSDA energies
for $J_1$, $J_2$, and $J_2'$.
The close agreement may be fortuitous
in view of the uncertainties mentioned above
and the approximations inherent in the SCAD method.
Nevertheless, we believe certain qualitative features of the SCAD results
are real: 1) The values for $J_1$ and $J_1'$ result primarily
from $j_d$, with relatively small contributions from fourth-order terms 
due to cancellation within $j_1$ and between $j_2$ and $j_3$.
2) The value for $J_2'$, the largest coupling, is dominated by a single term in
$j_1$, resulting from V overlap with the middle O 2p level.

For each set of four coupling constants,
we calculated the uniform susceptibility of the Hamiltonian
(\ref{eq:hamiltonian})
on periodic 20-spin clusters.
The susceptibility is given by:
\begin{equation}
\chi(T) = \frac{n (g \mu_B)^2}{N k_B T} \sum_{ij}\langle S_i^z S_j^z \rangle,
\label{eq:chi}
\end{equation}
where $n$ is the number of V atoms per gram and
$N$ is the number of sites in the cluster.
We take $g=1.67$ for all plots.
This was determined from the fit to the experimental magnetic susceptibility
data described
below.

To evaluate (\ref{eq:chi}), we calculate all eigenvalues
of the Hamiltonian---eigenvectors are not required.
We block-diagonalize the Hamiltonian with all possible symmetries:
translations, rotations, {\bf S}, and $S_z$ \cite{gros92a}.
The blocks are left with no degeneracies, so the eigenvalues
are calculated very efficiently using the Lanczos algorithm with
no reorthogonalization developed by Cullum and Willoughby\cite{cullum85a}.
This allows $\chi$ to be calculated exactly at all temperatures using one
Lanczos run for each symmetry sector.
The Hamiltonian for the 20-spin cluster has blocks as large as 36950.
Within each block at least the 400 lowest and
highest eigenvalues
are calculated, and an analytic density of states is assumed
for the middle eigenvalues.
This technique will be described
elsewhere.

\begin{figure}[tb]
\epsfxsize=3.0in\centerline{\epsffile{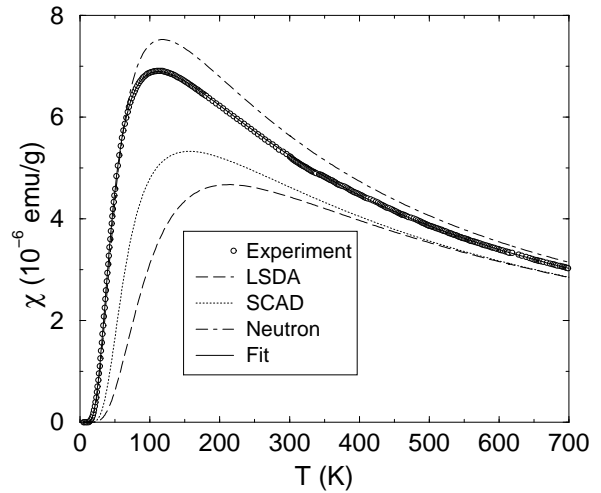}}
\caption{
Uniform magnetic susceptibilities calculated by exact diagonalization
of a 20-spin cluster.
The theoretical curves using the coupling constants
from Table \protect\ref{tab:allJs} are shown as lines,
while the circles show the experimental susceptibility of
Taniguchi, {\em et al.} \protect\cite{taniguchi95a}.
The theoretical fit to the susceptibility is the solid
curve that lies over the experimental points.
}
\label{fig:susc}
\end{figure}

The susceptibility of the full Hamiltonian (\ref{eq:hamiltonian})
calculated with each set of coupling constants in 
Table \ref{tab:allJs} is shown in Fig.\ \ref{fig:susc}.
The experimental susceptibility of Taniguchi, {\em et al.} \cite{taniguchi95a}
is shown for comparison.
All curves exhibit a spin gap, as evidenced by their low temperature
behavior, $\chi(T \rightarrow 0) \sim \exp (-\Delta/kT)$,
where $\Delta$ is the gap.
Both the LSDA and SCAD approaches overestimate the gap,
indicating that the calculated coupling constants are too large.
The coupling constants deduced from neutron scattering
are also shown \cite{fukumoto98a,weihong98a}.

Also shown in Fig.\ \ref{fig:susc}
is a curve generated using the coupling constants obtained from a 
least-squares fit of the susceptibility to the experimental results.
In the fitting procedure, we allow the $g$-value in eq.\ (\ref{eq:chi})
and all four $J$'s to vary.
At the best fit, we obtain the coupling constants listed as ``Fit'' in
Table \ref{tab:allJs} and shown as the line thicknesses in Fig.\ 1.
We find $g = 1.67$, which is smaller than the $g$-value
indicated by ESR measurements \cite{taniguchi97a}.
Near the minimum, the fitting function is quadratic.
The eigenvalues of the Hessian (scaled by an arbitrary constant) are 1,
0.046, 0.013, and 0.00039.
The smallness of the last eigenvalue indicates that in the
$\delta \{ J_1,  J_1',  J_2,  J_2' \} =
\{0.09,-0.57,0.81,-0.09\}$ direction from
the minimum, the least-squares fit is very soft.

The 20-spin cluster is sufficiently large compared with the
correlation length to describe the infinite system accurately.
The minimum triplet gap hardly varies between 20 and 32-spin
clusters: $\Delta_{20} = 9.92$ meV while $\Delta_{32} = 10.02$ meV
for the Fit Hamiltonian.

\begin{figure}[tbp]
\epsfxsize=3.0in\centerline{\epsffile{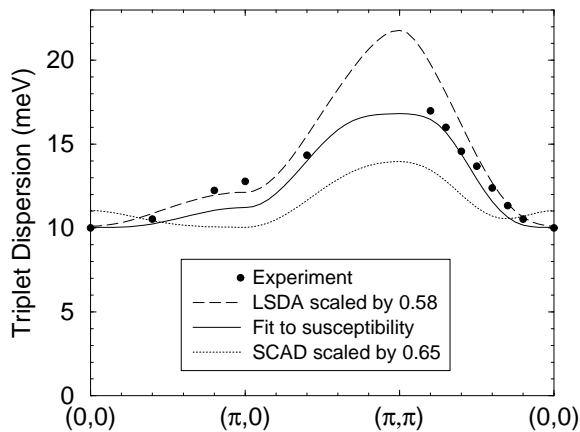}}
\vspace{.1in}
\caption{
The triplet dispersion $\Omega_Q$ in CaV$_4$O$_9$, calculated
from the fifth-order metaplaquette series expansion
of Weihong, Oitmaa, and Hamer \protect \cite{weihong98a}.
The circles are the neutron scattering data
of Kodama, {\em et al.} \protect\cite{kodama97a}.
The {\em ab initio} couplings have been rescaled so their minimum gaps
match the experimental minimum.
The solid line shows the dispersion of the (unrescaled) couplings determined
by fitting the experimental susceptibility.
}
\label{fig:triplet}
\end{figure}

Fig.\ \ref{fig:triplet} shows the triplet dispersion $\Omega_Q$
of the LSDA, SCAD, and susceptibility-fit coupling constants
calculated with the expansion in Ref.\ \cite{weihong98a}.
Since the LSDA and SCAD coupling constants overestimate the gap, we rescaled
their $J$'s by 0.58 and 0.65, respectively.
Both the Fit and rescaled LSDA $\Omega_Q$ agree
with the neutron scattering data reasonably well;
in particular, they correctly have minima at $Q = (0,0)$.

To conclude,
we have shown that the quantum-disordered phase in CaV$_4$O$_9$
can be predicted in {\em ab initio} fashion.
We calculated the coupling constants of the
Heisenberg Hamiltonian for CaV$_4$O$_9$ in two very different
first-principles approaches.
In both methods, the strongest coupling is found between
next-nearest-neighbor
V atoms on metaplaquettes---the weak coupling between nearest-neighbor V's
results from the cancellation among superexchange processes.
The uniform magnetic susceptibility for each set of
coupling constants is calculated using a novel finite-temperature
exact diagonalization technique, which shows
the Hamiltonians determined from both {\em ab initio} approaches
have quantum-disordered phases.
The Hamiltonian that best fits the experimental susceptibility is calculated,
and the agreement is remarkable.
Finally the triplet dispersion of the  {\em ab initio} and best
susceptibility-fit
Hamiltonians are shown to agree well with the neutron scattering data.

We thank Z. Weihong for the code to calculate
the curves in Fig.\ \ref{fig:triplet} and
N.E. Bonesteel,
J.L. Feldman,
R.E. Rudd,
M. Sato,
R.R.P. Singh,
and
C.C. Wan
for stimulating conversations.
This work was supported by the Office of Naval Research.
C.S.H was supported by the National Research Council, and W.E.P.
by NSF Grant DMR-9802076.
Computations were done at the Arctic Region Supercomputing Center and
at the DoD Major Shared Resource Centers at NAVOCEANO and CEWES.


\begin{references}



\bibitem{taniguchi95a}
S. Taniguchi {\it et~al.}, J. Phys.\ Soc.\ Japan {\bf 64},  2758  (1995).

\bibitem{kodama97a}
K. Kodama {\it et~al.}, J. Phys.\ Soc.\ Japan {\bf 66},  793  (1997).

\bibitem{gelfand96a}
M.~P. Gelfand {\it et~al.}, Phys.\ Rev.\ Lett.\ {\bf 77},  2794  (1996).

\bibitem{fukumoto98a}
Y. Fukumoto and A. Oguchi, J. Phys.\ Soc.\ Japan {\bf 67},  2205  (1998).

\bibitem{weihong98a}
Z. Weihong, J. Oitmaa, and C.~J. Hamer, Phys.\ Rev.\ B {\bf 58},  14147
  (1998).

\bibitem{theory}
N. Katoh and M. Imada, J. Phys.\ Soc.\ Japan {\bf 64},  4105  (1995);
K. Sano and K. Takano, {\em ibid.}           {\bf 65},  46  (1996);
K. Ueda, H. Koutani, M. Sigrist, and P.~A. Lee, Phys.\ Rev.\ Lett.\ {\bf 76},
  1932  (1996);
M. Troyer, H. Kontani, and K. Ueda, {\em ibid.}         {\bf 76},  3822 (1996);
O.~A. Starykh {\it et~al.}, {\em ibid.}         {\bf 77},  2558  (1996):
M. Albrecht, F. Mila, and D. Poilblanc, Phys.\ Rev.\ B {\bf 54},  15856 (1996);
Z. Weihong {\it et~al.}, {\em ibid.}    {\bf 55},  11377  (1997);
K. Takano and K. Sano, cond-mat  (9805153);
M.~A. Korotin  {\it et~al.}, cond-mat  (9901214).

\bibitem{bouloux73a}
J.-C. Bouloux and J. Galy, Acta Cryst.\ B {\bf 29},  1335  (1973).

\bibitem{pickett97a}
W.~E. Pickett, Phys.\ Rev.\ Lett.\ {\bf 79},  1746  (1997).

\bibitem{scad}
M.~J. Mehl, H.~T. Stokes, and L.~L. Boyer, J. Phys.\ Chem.\ Solids {\bf 57},
  1405  (1996);
L.~L. Boyer, H.~T. Stokes, and M.~J. Mehl, Ferroelectrics {\bf 194},  173
  (1997);
L.~L. Boyer, H.~T. Stokes, and M.~J. Mehl,  in {\em First Principles
  Calculations for Ferroelectrics}, AIP Conf.\ Proc.\, edited by R.~E. Cohen
  (AIP, Woodbury, NY, 1998), No.~436.

\bibitem{lapw}
An increased basis cutoff of E$_{max}$=22.6 Ry and a fixed
16 $k$-point mesh were used in the linearized augmented plane wave
calculations.

\bibitem{hedin71a}
L. Hedin and B.~I. Lundqvist, J. Phys.\ C {\bf 4},  2064  (1971).

\bibitem{lowdin50a}
P.~O. L{\"o}wdin, J. Chem.\ Phys.\ {\bf 18},  365  (1950).

\bibitem{gros92a}
C. Gros, Z.\ Phys.\ B {\bf 86},  359  (1992).

\bibitem{cullum85a}
J.~K. Cullum and R.~A. Willoughby, {\em Lanczos Algorithms
} (Birkhauser, Boston, 1985).

\bibitem{taniguchi97a}
S. Taniguchi {\it et~al.}, J. Phys.\ Soc.\ Japan {\bf 66},  3660  (1997).





\end{references}
\end{document}